\documentclass[aps,prl,amsmath,amssymb,reprint,groupedaddress,preprintnumbers]{revtex4}

\usepackage{enumitem}
\usepackage{graphicx}

\newcommand{\Tr}{\mathop{\mathrm{Tr}}\nolimits}
\newcommand{\sign}{\mathop{\mathrm{sign}}\nolimits}
\newcommand{\Z}{\mathcal{Z}}
\newcommand{\Op}{\mathcal{O}}
\newcommand{\cN}{\mathcal{N}}
\newcommand{\BS}{\,\mathbb{S}}
\newcommand{\z}[1]{\zeta_{#1}}
\newcommand{\M}{j}
\newcommand{\J}{\mathcal{J}}

\begin{document}

\title{Nonplanar Four-Loop Anomalous Dimensions of Twist-Two Operators in $\mathcal{N}=4$ Super Yang--Mills Theory: Higher Moment, General Result, and Cusp Anomalous Dimension}

\author{B.~A.~Kniehl}
\email[]{kniehl@desy.de}
\affiliation{Department of Physics, University of California at San Diego, 9500 Gilman Drive, La Jolla, CA 92093-0112, USA}
\affiliation{{II.} Institut f{\"u}r Theoretische Physik,
Universit{\"a}t Hamburg,
Luruper Chaussee 149, 22761 Hamburg, Germany}
\author{V.~N.~Velizhanin}
\email[]{vitaly.velizhanin@desy.de}
\affiliation{{II.} Institut f{\"u}r Theoretische Physik,
Universit{\"a}t Hamburg,
Luruper Chaussee 149, 22761 Hamburg, Germany}

\date{\today}

\begin{abstract}
  We consider the nonplanar universal anomalous dimension of twist-two operators at four loops in $\mathcal{N}=4$ supersymmetric Yang--Mills theory and push its direct diagrammatic calculation through Lorentz spin $j=20$, one unit beyond the state of the art, so as to confirm the correctness of the general, all-$j$ result conjectured previously by us \cite{Kniehl:2021ysp} imposing certain constraints on its analytic form.
Thanks to our new result, such constraints can be eliminated altogether.
By the same token, this allows us to re-derive, in a completely independent way, the nonplanar four-loop cusp anomalous dimension by taking the large-$j$ limit of the general result.  
\end{abstract}

\maketitle

The AdS/CFT correspondence \cite{Maldacena:1997re,Gubser:1998bc,Witten:1998qj},
also known as holographic duality, has been one of the most active and
tantalizing research topics in high-energy theory over the past quarter of a century. 
This implies that quantum gravity in anti--de Sitter space, with constant
negative curvature, is equivalent to a lower-dimensional nongravitational
quantum field theory of conformal type, $\cN=4$ supersymmetric Yang--Mills theory (SYM), living on the boundary of that gravitational space.
The AdS/CFT correspondence has led to a plethora of intriguing physical
insights and powerful novel methods of calculation.
In particular, integrability in the framework of the AdS/CFT correspondence has enabled the solution of the spectral problem for composite operators~\cite{Minahan:2002ve,Beisert:2003tq,Beisert:2003yb,Dolan:2003uh,Bena:2003wd,Kazakov:2004qf,Beisert:2004hm,Arutyunov:2004vx,Staudacher:2004tk,Beisert:2005di,Beisert:2005bm,Beisert:2005fw,Beisert:2005cw,Janik:2006dc,Hernandez:2006tk,Arutyunov:2006iu,Eden:2006rx,Beisert:2006ib,Beisert:2006ez,Beisert:2007hz,Beisert:2010jr,Arutyunov:2009zu,Gromov:2009tv,Arutyunov:2009ur,Bombardelli:2009ns,Gromov:2009bc,Arutyunov:2009ax,Gromov:2013pga,Gromov:2014caa}.

So far, investigations of the AdS/CFT correspondence have largely been confined
to the planar limit, in which Feynman diagrams of planar topologies contribute,
while nonplanar topologies are far more difficult to tackle and have, therefore, received much less attention.
In this connection, we recently considered the nonplanar universal anomalous dimension of twist-two operators at four loops in $\mathcal{N}=4$ SYM and conjectured for it a general expression, valid for arbitrary Lorentz spin $j$ \cite{Kniehl:2021ysp}.
This expression was reconstructed from the explicit results for fixed values of~$j$, through $j=18$, obtained via direct diagrammatic evaluation \cite{Kniehl:2020rip} with the help of modern tools of number theory, notably the \texttt{fplll} realization~\cite{fplll} of the Lenstra-Lenstra-Lovasz (LLL) algorithm \cite{Lenstra:1982}.
The step from $j\le18$ to arbitrary $j$ required for certain constraints to be imposed \cite{Kniehl:2021ysp}. 
Although we are quite confident in the correctness of our general expression, it is crucial for it to pass a final check to be on the safe side.
For this purpose, we calculate here the next, $j=20$, moment to find agreement with the corresponding value obtained from our general expression.

Specifically, we consider the following SU(4) singlet twist-two operators within $\mathcal{N}=4$ SYM:
\begin{eqnarray}
\mathcal{O}_{\mu _{1}\cdots\mu _{\M}}^{\lambda } &=&\hat{S}
\bar{\lambda}_{i}^{a}\gamma _{\mu _{1}}
    {\mathcal D}_{\mu _{2}}\cdots{\mathcal D}_{\mu _{\M}}\lambda ^{a\;i}\,,
    \nonumber
    \\
\mathcal{O}_{\mu _{1}\cdots\mu _{\M}}^{g} &=&\hat{S} G_{\rho \mu_{1}}^{a}{\mathcal
D}_{\mu _{2}} {\mathcal D}_{\mu _{3}}\cdots{\mathcal D}_{\mu _{\M-1}}G_{\phantom{a,\rho}\mu_{\M}}^{a,\rho}\,,
\nonumber\\
\mathcal{O}_{\mu _{1}\cdots\mu _{\M}}^{\phi } &=&\hat{S}
\bar{\phi}_{r}^{a}{\mathcal D}_{\mu _{1}} {\mathcal D}_{\mu _{2}}\cdots{\mathcal
D}_{\mu _{\M}}\phi _{r}^{a}\,,\label{phphs}
\end{eqnarray}
where ${\mathcal D}_{\mu_i}$ denotes the covariant derivative, the spinors $\lambda_{i}$ and the field strength tensor $G_{\rho \mu }$ describe fermions and gauge fields, respectively, and $\phi_{r}$ are the complex scalar fields appearing in ${\mathcal N}=4$ SYM.
The indices $i=1,2,3,4$ and $r=1,2,3$ refer to the SU(4) and SO(6)${}\simeq{}$SU(4) groups of inner symmetry, respectively.
The symbol $\hat{S}$ implies a symmetrization of the respective tensor in the Lorentz indices $\mu_{1},\ldots,\mu_{\M}$ and a subtraction of its traces. 
Certain combinations of the operators in Eq.~\eqref{phphs} are multiplicatively renormalized, namely, those in Eq.~\eqref{mrop3j} below.
Their anomalous dimensions are expressed through the so-called universal anomalous dimension,
\begin{equation}
  \gamma_{\mathrm {uni}}(\M) = \sum_{n=1}^{\infty}\gamma_{\mathrm {uni}}^{(n-1)}(\M)\,g^{2n}\,,
  \label{eq:gamuni}
\end{equation}
where $g^2=\lambda/(16\pi^2)$, $\lambda=g^2_{\mathrm {YM}}N_c$ is the 't~Hooft coupling constant, $g_{\mathrm {YM}}$ is the Yang--Mills coupling constant, and $N_c$ is the number of colors, up to integer argument shifts \cite{Kotikov:2002ab}.
In particular, $\gamma_{\mathrm{uni}}(j)$ is related to the anomalous dimension $\gamma_{\mathcal{O}^M_{\mathcal{Z}}}$ of the frequently studied twist-two operator 
\begin{equation}
\mathcal{O}^M_{\mathcal{Z}}=\Tr \mathcal{Z}{\mathcal D}_{\mu _{1}} {\mathcal D}_{\mu _{2}}\cdots{\mathcal
D}_{\mu _{M}}\mathcal{Z}\,,\label{OpZZ}
\end{equation}
where $\mathcal{Z}$ is one of the scalar fields $\phi_r$, which belong to the $\mathfrak{sl}(2)$ sub-sector of $\mathcal{N}=4$ SYM.
In fact, we have
\begin{equation}
\gamma_{\mathrm{uni}}(j-2)=\gamma_{\mathcal{O}^M_{\mathcal{Z}}}\,,
\end{equation}
which implies the shift of argument $j=M+2$.

\begin{figure}
\begin{center}
\includegraphics[width=0.47\textwidth]{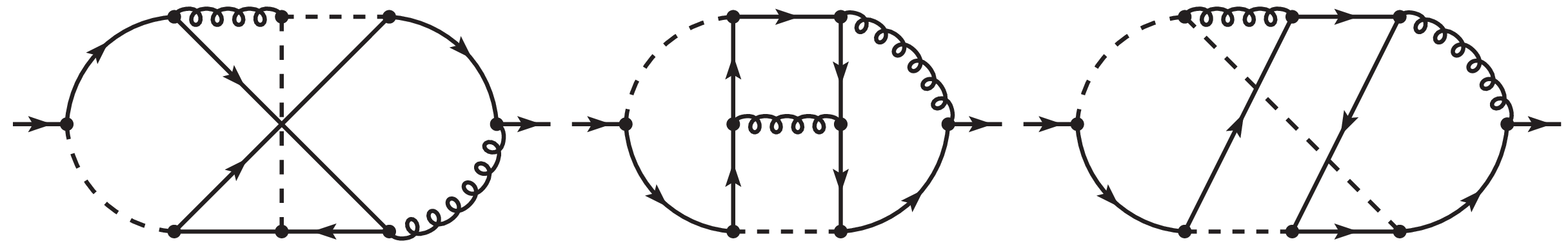}
\caption{Typical Feynman diagrams contributing to $\gamma_{\mathrm {uni},\mathrm{np}}^{(3)}(j)$.
The operators are inserted in the lines or gauge vertices.}
\label{fig:diagrams} 
\end{center}
\end{figure}

Nonplanar contributions arise from the Feynman diagrams that contain the following combination of color structures:
\begin{equation}
\quad
d^{abcd} =\frac{1}{6}\left[
\Tr\left(f^{paq}f^{qbr}f^{rcs}f^{sdp}\right)
 + \mathrm{five}\ bcd\ \mathrm{permutations}\right]\,,
\end{equation}
generating the color factor
\begin{equation}
  d_{44}=d^{abcd}d_{abcd}
=\frac{{N_c^4}}{32}\left(\frac{4}{3}+\frac{48}{N_c^2}\right)\,.
\end{equation}
Therefore, the nonplanar (np) four-loop ($n=4$) contributions of Lorentz spin $j$ to the universal anomalous dimension in Eq.~\eqref{eq:gamuni}, $\gamma_{\mathrm{uni,np}}^{(3)}(j)g^8$, carry the common factor $48/N_c^2$.
Typical Feynman diagrams contributing to $\gamma_{\mathrm{uni,np}}^{(3)}(j)$ are depicted in Fig.~\ref{fig:diagrams}.

In principle, we can calculate the nonplanar contribution to the universal anomalous dimension for any operator (or multiplicatively renormalized combination of operators) in the theory, for example, for the operator $\Op_{\Z}^M$ in Eq.~(\ref{OpZZ}), as we did in Ref.~\cite{Kniehl:2023bbk}.
However, as explained in Refs.~\cite{Kniehl:2020rip,Kniehl:2021ysp}, the calculation may be considerably simplified in the case at hand by exploiting the fact that the nonplanar contribution appears for the first time at four-loop order and does not require any renormalization.

This simplification may be easily explained by means of well-known results at leading order.
The matrix of anomalous dimensions for the operators in Eq.~\eqref{phphs}, sandwiched between definite states (fermions, gauge bosons, and scalars), has the following elements:
\begin{eqnarray}
\gamma^{(0)}_{{gg}}&=&-4S_1(j)+\frac{4}{j-1}-\frac{4}{j}+\frac{4}{j+1}-\frac{4}{j+2}\,,\qquad
\gamma^{(0)}_{{\lambda g}} = \frac{8}{j}-\frac{16}{j+1}+\frac{16}{j+2}\,,
\nonumber \\
\gamma^{(0)}_{{\phi g}}&=&\frac{12}{j+1}-\frac{12}{j+2}\,,\quad
\gamma^{(0)}_{{g\lambda}} = \frac{4}{j-1}-\frac{4}{j}+\frac{2}{j+1}\,,\quad
\gamma^{(0)}_{{\lambda\phi}} = \frac{8}{j}\,,\quad
\gamma^{(0)}_{{\phi\lambda}} = \frac{6}{j+1}\,,
\nonumber \\
\gamma^{(0)}_{{\lambda\lambda}} &=&-4S_1(j)+\frac{8}{j}-\frac{8}{j+1}\,,\qquad
\gamma^{(0)}_{{g\phi}} = \frac{4}{j-1}-\frac{4}{j}\,,\qquad
\gamma^{(0)}_{{\phi \phi}} = -4 S_1(j)\,,
\label{LOAD}
\end{eqnarray}
where $S_1(j)=\sum_{i=1}^{j}\frac{1}{i}$ is the simplest harmonic sum.
The matrix that diagonalizes the matrix of anomalous dimensions in Eq.~\eqref{LOAD} can be used to construct the following multiplicatively renormalizable operators:
\begin{eqnarray}
\Op^{T_j}_{\mu_1\ldots\mu_j} & = & \Op^g_{\mu_1\ldots\mu_j} +
\Op^\lambda_{\mu_1\ldots\mu_j} + \Op^\phi_{\mu_1\ldots\mu_j}\,,
\nonumber\\
\Op^{\Sigma_j}_{\mu_1\ldots\mu_j} & = & - 2(j-1)\Op^g_{\mu_1\ldots\mu_j} +
\Op^\lambda_{\mu_1\ldots\mu_j} + \frac{2}{3}(j+1)\Op^\phi_{\mu_1\ldots\mu_j}\,,
\nonumber\\
\Op^{\Xi_j}_{\mu_1\ldots\mu_j} & = & -\frac{j-1}{j+2}\Op^g_{\mu_1\ldots\mu_j}
+ \Op^\lambda_{\mu_1\ldots\mu_j} -
\frac{j+1}{j}\Op^\phi_{\mu_1\ldots\mu_j}\,,
\label{mrop3j}
\end{eqnarray}
whose anomalous dimensions are $\gamma_{\mathrm {uni}}^{(0)}(j)$, $\gamma_{\mathrm {uni}}^{(0)}(j+2)$, and $\gamma_{\mathrm {uni}}^{(0)}(j+4)$, respectively, where
\begin{equation}
  \gamma_{\mathrm {uni}}^{(0)}(j)=-4S_1(j-2)=-4S_1(M)\,.
  \label{gammauni0}
\end{equation}

Sandwiching the operators in Eq.~\eqref{mrop3j} between different states, we obtain the following set of relations:
\begin{eqnarray}
\gamma_{g\lambda}^{(0)}+\gamma_{\lambda\lambda}^{(0)}+\gamma_{\phi\lambda}^{(0)} &=& \gamma_{\mathrm {uni}}^{(0)}(j)\,,\nonumber\\
-2(j-1)\gamma_{g\lambda}^{(0)}+                 \gamma_{\lambda\lambda}^{(0)}+\frac{2}{3}(j+1)          \gamma_{\phi\lambda}^{(0)} &=& \gamma_{\mathrm {uni}}^{(0)}(j+2)\,,\nonumber\\
-\frac{j-1}{j+2}\gamma_{g\lambda}^{(0)}+          \gamma_{\lambda\lambda}^{(0)}-\frac{j+1}{j}               \gamma_{\phi\lambda}^{(0)} &=& \gamma_{\mathrm {uni}}^{(0)}(j+4)\,,\label{Op3l}
\end{eqnarray}
where, on the left-hand sides, diagonal elements are normalized to unity and the arguments of the matrix elements are equal to $j$. 
If we calculate the anomalous dimension for the operators in Eq.~\eqref{phphs} at some fixed value of $j$, we thus obtain the result for the universal anomalous dimension not only for $j$, but also for $(j+2)$ and $(j+4)$.

Notice that Eqs.~\eqref{mrop3j} and \eqref{Op3l} carry over from the planar one-loop case to the nonplanar four-loop one, as the nonplanar contribution appears for the first time at four-loop order and so does not require any renormalization. In order to obtain $\gamma_{\mathrm {uni}}^{(3)}(20)$, as in our case, it is sufficient to compute, for example, $\gamma_{\phi\lambda}^{(3)}(16)$.
Moreover, the calculation of a matrix element like $\gamma_{\phi\lambda}^{(3)}(16)$ does not involve the operator vertex with four gauge lines, the Feynman rules for which contain a huge number of terms.
In a similar way, we managed to reach $j=18$ in Ref.~\cite{Kniehl:2020rip}.

Our calculations are performed using modern tools of computer algebra that are publicly available in combination with custom-made scripts of symbolic manipulation, the computational setup being similar to Ref.~\cite{Kniehl:2020rip}.
Specifically, we generate the Feynman diagrams using the program package \texttt{DIANA}~\cite{Tentyukov:1999is}, which calls \texttt{QGRAF}~\cite{Nogueira:1991ex}, take the color traces using the \texttt{FORM}~\cite{Vermaseren:2000nd} program package \texttt{COLOR}~\cite{vanRitbergen:1998pn}, and evaluate the resulting expressions using the \texttt{FORM}~\cite{Vermaseren:2000nd} program package \texttt{FORCER} \cite{Ruijl:2017cxj}, which was developed for this type of computations and was also used to compute anomalous dimensions of twist-two operators in QCD~\cite{Moch:2017uml,Moch:2018wjh}. 

After a lengthy computation, we find the central ingredient to be
\begin{equation}
\frac{N_c^2}{48}\,\gamma_{\phi\lambda,\mathrm{np}}^{(3)}(16) =
-\frac{15321193161485705123647}{736310761772826240000}
-\frac{224810362251649}{3863315055600}\,\z3 
+\frac{248121611}{2603601}\,\z5\,.
\end{equation}
Substituting this expression into the relation
\begin{equation}
\gamma_{\mathrm{uni}}(20)=
-\frac{385}{144}\,\gamma_{\phi\lambda}(16) 
+ \frac{175}{186}\,\gamma_{\mathrm{uni}}(16)
+\frac{11}{186}\,\gamma_{\mathrm{uni}}(18)\,,
\end{equation}
which follows from the system of equations in Eq.~(\ref{Op3l}) for $j=16$, we
find
\begin{eqnarray}
\frac{N_c^2}{48}\,\gamma_{\mathrm{uni},\mathrm{np}}^{(3)}(20)&=&
\frac{1399928337444356711866218313}{4337539760261740032000000}
+ \frac{27787856664398189}{28974862917000}\,\z3\nonumber\\
&&{}-\frac{203755669038601}{104248184040}\,\z5\,.
\label{unij20}
\end{eqnarray}
This exactly agrees with our general expression \cite{Kniehl:2021ysp}, 
\begin{eqnarray}
\frac{N_c^2}{48}\,\gamma_{\mathrm {uni},\mathrm {np}}^{(3)}(j)&=& 
4\left(
2\BS_{5,2}
-2\BS_{4,3}
-4\BS_{1,2,4}
-2\BS_{1,3,3}
+2\BS_{1,4,2}
+4\BS_{1,5,1}
-6\BS_{2,2,3}\right.
\nonumber\\
&&{}+2\BS_{2,3,2}
+4\BS_{2,4,1}
+12\BS_{3,1,3}
-6\BS_{3,2,2}
-2\BS_{3,3,1}
+2\BS_{4,1,2}
+6\BS_{4,2,1}
\nonumber\\
&&{}
-12\BS_{5,1,1}
+8\BS_{1,1,1,4}
+4\BS_{1,1,2,3}
-8\BS_{1,1,3,2}
-4\BS_{1,1,4,1}
-\BS_{1,2,1,3}
+3\BS_{1,2,2,2}
\nonumber\\
&&{}
+6\BS_{1,2,3,1}
-14\BS_{1,3,1,2}
-3\BS_{1,3,2,1}
+9\BS_{1,4,1,1}
-3\BS_{2,1,1,3}
-\BS_{2,1,2,2}
+\BS_{2,2,2,1}
\nonumber\\
&&{}
+3\BS_{2,3,1,1}
-8\BS_{3,1,1,2}
-8\BS_{3,1,2,1}
+12\BS_{3,2,1,1}
+4\BS_{4,1,1,1}
+4\BS_{1,1,1,2,2}
\nonumber\\
&&{}
-12\BS_{1,1,1,3,1}
+4\BS_{1,1,2,1,2}
-4\BS_{1,1,2,2,1}
+8\BS_{1,1,3,1,1}
+8\BS_{1,2,1,1,2}
+2\BS_{1,2,1,2,1}
\nonumber\\
&&{}
-\left.14\BS_{1,2,2,1,1}
+4\BS_{1,3,1,1,1}
+4\BS_{2,1,1,1,2}
+4\BS_{2,1,1,2,1}
-4\BS_{2,1,2,1,1}
-4\BS_{2,2,1,1,1}
\right)
\nonumber\\
&&{}
+8\z3\left(8\BS_4-9 \BS_{1,3}-3 \BS_{2,2}-4 \BS_{3,1}+4 \BS_{1,1,2}+5 \BS_{1,2,1}-\BS_{2,1,1}\right)
\nonumber\\
&&{}
-40\z5\BS_{1}^2\,,
\label{npadres}
\end{eqnarray}
for $j=20$, where $\BS_{\vec{\mathbf{a}}}=\BS_{\vec{\mathbf{a}}}(j-2)=\BS_{\vec{\mathbf{a}}}(M)$ for the ease of notation, and the binomial and nested harmonic sums are defined as~\cite{Vermaseren:1998uu}
\begin{eqnarray} 
\BS_{a_1,\ldots,a_n}(M)&=&(-1)^M\sum_{j=1}^{M}(-1)^j\binom{M}{j}\binom{M+j}{j}S_{a_1,...,a_n}(j)\,,
\label{BinomialSums}
\\
S_{a_1,\ldots,a_n}(M)&=&\sum^{M}_{j=1} \frac{[\sign(a_1)]^{j}}{j^{\vert a_1\vert}}
\,S_{a_2,\ldots,a_n}(j)\, ,\qquad
S_a (M)=\sum^{M}_{j=1} \frac{[\sign(a)]^{j}}{j^{\vert a\vert}}\,.
\label{vhs}
\end{eqnarray}

We recall that, in Ref.~\cite{Kniehl:2021ysp}, the all-$j$ result in Eq.~\eqref{npadres} was conjectured from the knowledge of the first 8 nonvanishing moments
$\gamma_{\mathrm{uni},\mathrm{np}}^{(3)}(j)$, with $=4,6,\ldots,18$, assuming 14 additional constraints in the form of expected analytic properties of its rational part, namely, 8 from the large-$j$ limit, 5 from the  Balitsky--Fadin--Kuraev--Lipatov (BFKL)~\cite{Lipatov:1976zz,Kuraev:1977fs,Balitsky:1978ic} equation through next-to-leading-logarithmic approximation \cite{Kotikov:2002ab,Fadin:1998py,Kotikov:2000pm}, and 1 from the double-logarithmic equation~\cite{Kirschner:1982qf,Kirschner:1983di}.
In particular, we injected the knowledge of the four-loop cusp anomalous dimension $\gamma_{\mathrm{cusp}}^{(3)}=2\Gamma_4^{\mathcal{N}=4}$ \cite{Henn:2019swt,Huber:2019fxe,vonManteuffel:2020vjv} to fix the coefficient of the leading large-$j$ asymptotic term, proportional to $\ln j$, via the relation
\begin{equation}
\gamma_{\mathrm{uni}}^{(3)}(j)\overset{j\to\infty}{=}
\gamma_{\mathrm{cusp}}^{(3)}\ln j+\mathcal{O}(1)\,.
\label{eq:cusp}
\end{equation}
This actually accounts for 2 of the 8 constraints from the large-$j$ limit mentioned above, by fixing the coefficients of $\z3^2\ln j$ and $\z6\ln j$ in $\gamma_{\mathrm{uni},\mathrm{np}}^{(3)}(j)$ (see Eq.~\eqref{largejexp} below).
For lack of space, we refer to Ref.~\cite{Kniehl:2021ysp} for full details on the remaining 12 constraints.

Our direct diagrammatic derivation of the ninth nonvanishing moment $\gamma_{\mathrm{uni},\mathrm{np}}^{(3)}(20)$, given in Eq.~\eqref{unij20}, totally changes the game.
Owing to the highly nonlinear response property of the LLL algorithm \cite{Lenstra:1982}, we are now in a position to do without any additional assumptions, {\it i.e.}, to drop all of the above-mentioned 14 constraints in one sweep.
By the same token, this implies that we independently establish the latter.
In particular, we provide a completely independent derivation of the nonplanar four-loop cusp anomalous dimension $\gamma_{\mathrm{cusp},\mathrm{np}}^{(3)}(j)$, by taking the large-$j$ limit of Eq.~\eqref{npadres}.
This expansion is also of interest from the string theory point of view, even beyond the leading term, which further motivates us to consider it here.
This task is similar to the large-$u$ expansion of the $\eta_{\vec{\bf{a}}}$ function that appears within the quantum spectral curve framework, and we may apply an appropriately modified version of the \texttt{MATHEMATICA} code originally developed for this purpose in Ref.~\cite{Velizhanin:2021bdh}, which we also successfully tested for the planar part of $\gamma_{\mathrm{uni}}^{(3)}(j)$. 
We find
\begin{eqnarray}
&&\frac{N_c^2}{48}\,\gamma_{\mathrm{uni},\mathrm{np}}^{(3)}(j)\overset{j\to\infty}{=}
- 2 \left(12 \z3^2 + 31 \z6\right) \ln j
+4 \z3 \z4 - 20 \z2 \z5 - 175 \z7
-\frac{1}{j}\left(12 \z3^2 + 31 \z6\right)
+\frac{1}{j^2}\left[- 32 \left(\z2 + 2 \z3\right) \ln^2j
\vphantom{\frac{31}{6}}\right.
\nonumber\\
&&{}+\left.
8  \left(4 \z2 + \z3 - 8 \z4\right) \ln j
+ 32 \z2 + 24 \z3 + 84 \z4 + 24 \z2 \z3 - 20 \z5  + 2 \z3^2 +
\frac{31}{6} \z6
\right]
+\mathcal{O}\left(\frac{1}{j^2}\right)\,.
\label{largejexp}
\end{eqnarray}
The result for $\gamma_{\mathrm{cusp},\mathrm{np}}^{(3)}$ that we read off from Eq.~\eqref{largejexp} according to Eq.~\eqref{eq:cusp} agrees with that in Refs.~\cite{Henn:2019swt,Huber:2019fxe,vonManteuffel:2020vjv}, providing an independent confirmation of the latter.
It is interesting to observe that the $1/j$ term in Eq.~\eqref{largejexp} does not contain a term proportional to $\ln j$ and equals $\gamma_{\mathrm{cusp},\mathrm{np}}^{(3)}/2$.
Notice also that, because our general result~(\ref{npadres}) is expressed through reciprocity-respecting binomial harmonic sums, its large-$j$ expansion in Eq.~\eqref{largejexp} only contains even negative powers of $\J=\sqrt{j(j+1)}$, according to the general consideration in Ref.~\cite{Basso:2006nk} based on reciprocity \cite{Dokshitzer:2005bf,Dokshitzer:2006nm}.
Specifically, the large-$\J$ expansion of Eq.~\eqref{npadres} reads
\begin{eqnarray}
&&\frac{N_c^2}{48}\,\gamma_{\mathrm{uni},\mathrm{np}}^{(3)}(j)\overset{\J\to\infty}{=}
\gamma_{\mathrm{cusp},\mathrm{np}}^{(3)}\ln{\J} 
+ 4 \z3 \z4 - 20 \z2 \z5 - 175 \z7
+\frac{1}{\J^2}
\left[
-32   (\z2 +2 \z3)\ln^2{\J} 
+ 8  (4 \z2 + \z3 - 8 \z4)\ln{\J} 
\vphantom{\frac{31}{3}}\right.
\nonumber\\
&&{}+\left.
32 \z2 
+ 24 \z3 
+ 84 \z4 
+ 24 \z2 \z3 
- 20 \z5 
- 4 \z3^2 
- \frac{31}{3} \z6
\right]
+\mathcal{O}\left(\frac{1}{\J^4}\right)\,.
\label{largeJexp}
\end{eqnarray}
The simplicity of the term proportional to $1/j$ in Eq.~(\ref{largejexp}) and its relationship to the leading one may be understood by observing that they both originate from the term proportional to $\ln\J$ in Eq.~(\ref{largeJexp}).
We emphasize that the only way to find subleading terms of the expansions in Eqs.~(\ref{largejexp}) and (\ref{largeJexp}) is to first establish the all-$j$ result in Eq.~(\ref{npadres}) as we did.

To summarize, using advanced computer algebraic techniques of Feynman diagram calculus, we have deepened our knowledge of the nonplanar sector of $\cN=4$ SYM by studying the universal anomalous dimension of the local, gauge-invariant, SU(4)-singlet, twist-two operators of definite Lorentz spin $j$ at four loops.
Specifically, we have pushed the state of the art by one unit, to $j=20$.
Our new result confirms the all-$j$ result previously conjectured by us \cite{Kniehl:2021ysp} imposing certain constraints on the analytic form of its rational part.
In particular, this included the large-$j$ asymptotic limit as encoded in the cusp anomalous dimension, which was derived in Refs.~\cite{Henn:2019swt,Huber:2019fxe,vonManteuffel:2020vjv} using completely different techniques, based on Wilson loops and form factors.
By our direct diagrammatic analysis at $j=20$, we managed to cross the threshold beyond which the LLL algorithm \cite{Lenstra:1982} is able to uniquely determine the full all-$j$ result without any additional assumptions, thus featuring its highly nonlinear response property.
In other words, we were put in a position to independently validate in one go the 14 assumptions that were indispensable to make the LLL algorithm run successfully in Ref.~\cite{Kniehl:2021ysp}.
Most importantly, we so re-derived, in a completely independent way, the nonplanar four-loop cusp anomalous dimension in $\cN=4$ SYM \cite{Henn:2019swt,Huber:2019fxe,vonManteuffel:2020vjv}.

To encounter the turning point for the all-$j$ reconstruction at a $j$ value as low as 20 came as a surprise and underlines how enormously powerful our new LLL-based method is.
This promising observation within $\cN=4$ SYM is bound to carry over to real QCD.
In fact, this strongly motivates us to push the diagrammatic evaluation of anomalous-dimension moments in QCD to higher Lorentz spin, albeit this is an enormously tedious task, because the prospect of reaching the turning point for the all-$j$ reconstruction with available computer resources now appears much more realistic.
In turn, this will allow us to speed up the exploration of scaling violations in parton distributions, fragmentation functions and related objects in QCD to high precision, at multi-loop order in perturbation theory.
In fact, substantial numerical uncertainties due to extrapolations based on limited sets of low-$j$ moments, which have so far been inevitable (see, e.g., Figs.~5 and 6 in Ref.~\cite{Moch:2017uml}), can thus be quenched.
We so expect our achievement to have a long-term high impact on the exploitation of the physics potential of present and future colliders involving hadron beams or fixed targets, such as the CERN Large Hadron Collider and the BNL Electron Ion Collider.


We thank A.A. Tseytlin for motivating us to derive Eq.~\eqref{largejexp} for its usefulness for dedicated studies in string theory, and I.V. Surnin for sharing with us a method previously applied in Ref.~\cite{Velizhanin:2021bdh}.
The work of B.A.K. was supported by the German Research Foundation DFG through Grant No.~KN~365/16-1.
The work of V.N.V. was supported by the Russian Science Foundation under Grant No.~23-22-00311.


\begin{thebibliography}{99}

\bibitem{Kniehl:2021ysp}
B.~A.~Kniehl and V.~N.~Velizhanin,
Non-planar universal anomalous dimension of twist-two operators with general Lorentz spin at four loops in $\mathcal{N}=4$ SYM theory,
Nucl.\ Phys.\ B \textbf{968} (2021) 115429
[arXiv:2103.16420 [hep-th]].

\bibitem{Maldacena:1997re} 
  J.~Maldacena,
  The Large-$N$ Limit of Superconformal Field Theories and Supergravity,
  Int.\ J.\ Theor.\ Phys.\  {\bf 38} (1999) 1113--1133
  [Adv.\ Theor.\ Math.\ Phys.\  {\bf 2} (1998) 231--252] 
  [hep-th/9711200].

\bibitem{Gubser:1998bc} 
  S.~S.~Gubser, I.~R.~Klebanov, and A.~M.~Polyakov,
  Gauge theory correlators from non-critical string theory,
  Phys.\ Lett.\ B {\bf 428} (1998) 105--114
  [hep-th/9802109].

\bibitem{Witten:1998qj} 
  E.~Witten,
  Anti de Sitter Space and Holography,
  Adv.\ Theor.\ Math.\ Phys.\  {\bf 2} (1998) 253--291
  [hep-th/9802150].

\bibitem{Minahan:2002ve} 
  J.~A.~Minahan and K.~Zarembo,
  The Bethe-ansatz for $\mathcal{N}=4$ super Yang-Mills,
  J. High Energy Phys.\ {\bf 03} (2003) 013 
  [hep-th/0212208].

\bibitem{Beisert:2003tq}
N.~Beisert, C.~Kristjansen, and M.~Staudacher,
The dilatation operator of conformal $\mathcal{N}=4$ super-Yang--Mills theory,
Nucl.\ Phys.\ B \textbf{664} (2003) 131--184
[arXiv:hep-th/0303060 [hep-th]].

\bibitem{Beisert:2003yb}
N.~Beisert and M.~Staudacher,
The $\mathcal{N}=4$ SYM integrable super spin chain,
Nucl.\ Phys.\ B \textbf{670} (2003) 439--463
[arXiv:hep-th/0307042 [hep-th]].

\bibitem{Dolan:2003uh}
L.~Dolan, C.~R.~Nappi, and E.~Witten,
A Relation between approaches to integrability in superconformal Yang-Mills theory,
J. High Energy Phys.\ \textbf{10} (2003) 017
[arXiv:hep-th/0308089 [hep-th]].

\bibitem{Bena:2003wd}
I.~Bena, J.~Polchinski, and R.~Roiban,
Hidden symmetries of the AdS$_5\times{}$S$^5$ superstring,
Phys. Rev. D \textbf{69} (2004) 046002
[arXiv:hep-th/0305116 [hep-th]].

\bibitem{Kazakov:2004qf}
V.~A.~Kazakov, A.~Marshakov, J.~A.~Minahan, and K.~Zarembo,
Classical/quantum integrability in AdS/CFT,
J. High Energy Phys.\ \textbf{05} (2004) 024
[arXiv:hep-th/0402207 [hep-th]].

\bibitem{Beisert:2004hm}
N.~Beisert, V.~Dippel, and M.~Staudacher,
A novel long-range spin chain and planar $\mathcal{N}=4$ super Yang-Mills,
J. High Energy Phys.\ \textbf{07} (2004) 075
[arXiv:hep-th/0405001 [hep-th]].

\bibitem{Arutyunov:2004vx}
G.~Arutyunov, S.~Frolov, and M.~Staudacher,
Bethe ansatz for quantum strings,
J. High Energy Phys.\ \textbf{10} (2004) 016
[arXiv:hep-th/0406256 [hep-th]].

\bibitem{Staudacher:2004tk}
M.~Staudacher,
The factorized S-matrix of CFT/AdS,
J. High Energy Phys.\ \textbf{05} (2005) 054
[arXiv:hep-th/0412188 [hep-th]].

\bibitem{Beisert:2005di}
N.~Beisert, V.~A.~Kazakov, K.~Sakai, and K.~Zarembo,
Complete spectrum of long operators in $\mathcal{N}=4$ SYM at one loop,
J. High Energy Phys.\ \textbf{07} (2005) 030
[arXiv:hep-th/0503200 [hep-th]].

\bibitem{Beisert:2005bm}
N.~Beisert, V.~A.~Kazakov, K.~Sakai, and K.~Zarembo,
The Algebraic Curve of Classical Superstrings on $AdS_5\times S^5$,
Commun.\ Math.\ Phys.\ \textbf{263} (2006) 659--710
[arXiv:hep-th/0502226 [hep-th]].

\bibitem{Beisert:2005fw} 
  N.~Beisert and M.~Staudacher,
  Long-range $\mathfrak{psu}(2,2|4)$ Bethe ans\"atze for gauge theory and strings,
  Nucl.\ Phys.\ B {\bf 727} (2005) 1--62
  [hep-th/0504190].

\bibitem{Beisert:2005cw}
N.~Beisert and A.~A.~Tseytlin,
On quantum corrections to spinning strings and Bethe equations,
Phys.\ Lett.\ B \textbf{629} (2005) 102--110
[arXiv:hep-th/0509084 [hep-th]].

\bibitem{Janik:2006dc}
R.~A.~Janik,
The AdS$_5\times S^5$ superstring worldsheet $S$ matrix and crossing symmetry,
Phys. Rev. D \textbf{73} (2006) 086006
[arXiv:hep-th/0603038 [hep-th]].

\bibitem{Hernandez:2006tk}
R.~Hern\'andez and E.~L\'opez,
Quantum corrections to the string Bethe ansatz,
J. High Energy Phys.\ \textbf{07} (2006) 004
[arXiv:hep-th/0603204 [hep-th]].

\bibitem{Arutyunov:2006iu}
G.~Arutyunov and S.~Frolov,
On AdS$_5\times{}$S$^5$ string S-matrix,
Phys. Lett. B \textbf{639} (2006) 378--382
[arXiv:hep-th/0604043 [hep-th]].

\bibitem{Eden:2006rx}
B.~Eden and M.~Staudacher,
Integrability and transcendentality,
J. Stat. Mech. \textbf{0611} (2006) P11014
[arXiv:hep-th/0603157 [hep-th]].

\bibitem{Beisert:2006ib}
N.~Beisert, R.~Hern\'andez, and E.~L\'opez,
A crossing-symmetric phase for $AdS_5\times S^5$ strings,
J. High Energy Phys.\ \textbf{11} (2006) 070
[arXiv:hep-th/0609044 [hep-th]].

\bibitem{Beisert:2006ez} 
  N.~Beisert, B.~Eden, and M.~Staudacher,
  Transcendentality and crossing,
  J.\ Stat.\ Mech.\ ({\bf 2007}) P01021 
  [hep-th/0610251].

\bibitem{Beisert:2007hz}
N.~Beisert, T.~McLoughlin, and R.~Roiban,
Four-loop dressing phase of $\mathcal{N}=4$ super-Yang-Mills theory,
Phys. Rev. D \textbf{76} (2007) 046002
[arXiv:0705.0321 [hep-th]].

\bibitem{Beisert:2010jr} 
  N.~Beisert {\it et al.},
  Review of AdS/CFT Integrability: An Overview,
  Lett.\ Math.\ Phys.\  {\bf 99} (2012) 3--32
  [arXiv:1012.3982 [hep-th]].

\bibitem{Arutyunov:2009zu}
G.~Arutyunov and S.~Frolov,
String hypothesis for the AdS$_5\times{}$S$^5$ mirror,
J. High Energy Phys.\ \textbf{03} (2009) 152
[arXiv:0901.1417 [hep-th]].

\bibitem{Gromov:2009tv}
N.~Gromov, V.~Kazakov, and P.~Vieira,
Exact Spectrum of Anomalous Dimensions of Planar $\mathcal{N}=4$ Supersymmetric Yang-Mills Theory,
Phys.\ Rev.\ Lett.\ \textbf{103} (2009) 131601
[arXiv:0901.3753 [hep-th]].

\bibitem{Arutyunov:2009ur}
G.~Arutyunov and S.~Frolov,
Thermodynamic Bethe ansatz for the AdS$_5\times{}$S$^5$ mirror model,
J. High Energy Phys.\ \textbf{05} (2009) 068
[arXiv:0903.0141 [hep-th]].

\bibitem{Bombardelli:2009ns}
D.~Bombardelli, D.~Fioravanti, and R.~Tateo,
Thermodynamic Bethe ansatz for planar AdS/CFT: a proposal,
J. Phys.\ A: Math.\ Theor.\ \textbf{42} (2009) 375401
[arXiv:0902.3930 [hep-th]].

\bibitem{Gromov:2009bc}
N.~Gromov, V.~Kazakov, A.~Kozak, and P.~Vieira,
Exact Spectrum of Anomalous Dimensions of Planar $N = 4$ Supersymmetric Yang--Mills Theory: TBA and excited states,
Lett. Math. Phys. \textbf{91} (2010) 265--287
[arXiv:0902.4458 [hep-th]].

\bibitem{Arutyunov:2009ax}
G.~Arutyunov, S.~Frolov, and R.~Suzuki,
Exploring the mirror TBA,
J. High Energy Phys.\ \textbf{05} (2010) 031
[arXiv:0911.2224 [hep-th]].

\bibitem{Gromov:2013pga} 
  N.~Gromov, V.~Kazakov, S.~Leurent, and D.~Volin,
  Quantum Spectral Curve for Planar $\mathcal{N} = 4$ Super-Yang-Mills Theory,
  Phys.\ Rev.\ Lett.\  {\bf 112} (2014) 011602 
  [arXiv:1305.1939 [hep-th]].

\bibitem{Gromov:2014caa} 
  N.~Gromov, V.~Kazakov, S.~Leurent, and D.~Volin,
  Quantum spectral curve for arbitrary state/operator in AdS$_{5}$/CFT$_{4}$,
  J. High Energy Phys.\ {\bf 09} (2015) 187 
  [arXiv:1405.4857 [hep-th]].

\bibitem{Kniehl:2020rip} 
  B.~A.~Kniehl and V.~N.~Velizhanin,
  Nonplanar Cusp and Transcendental Anomalous Dimension at Four Loops in $\mathcal{N}=4$ Supersymmetric Yang-Mills Theory,
  Phys.\ Rev.\ Lett.\  {\bf 126} (2021) 061603
  [arXiv:2010.13772 [hep-th]].

\bibitem{fplll}
  M.~Albrech, D.~Cad\'{e}, X.~Pujol, and D.~Stehl\'{e} (The {FPLLL} development team),
\texttt{fplll}, a lattice reduction library, 2016,
available at \url{https://github.com/fplll/fplll}.

\bibitem{Lenstra:1982} 
  A.~K.~Lenstra, H.~W.~Lenstra, and L.~Lov\'asz,
  Factoring Polynomials with Rational Coefficients,
  Math.\ Ann.\ {\bf 261} (1982) 515--534.

\bibitem{Kotikov:2002ab} 
  A.~V.~Kotikov and L.~N.~Lipatov,
  DGLAP and BFKL equations in the $N=4$ supersymmetric gauge theory,
  Nucl.\ Phys.\ B {\bf 661} (2003) 19--61;
  erratum: Nucl.\ Phys.\ B {\bf 685} (2004) 405--407
  [hep-ph/0208220].

\bibitem{Kniehl:2023bbk}
B.~A.~Kniehl and V.~N.~Velizhanin,
Anomalous dimensions of twist-two operators in extended $\mathcal{N}=2$ and $\mathcal{N}=4$ super Yang-Mills theories,
Nucl. Phys. B \textbf{1001} (2024) 116511
[arXiv:2312.05888 [hep-th]].

\bibitem{Tentyukov:1999is} 
  M.~Tentyukov and J.~Fleischer,
  A Feynman diagram analyzer DIANA,
  Comput.\ Phys.\ Commun.\  {\bf 132} (2000) 124--141
  (2000)
  [hep-ph/9904258].

\bibitem{Nogueira:1991ex} 
  P.~Nogueira,
  Automatic Feynman Graph Generation,
  J.\ Comput.\ Phys.\  {\bf 105} (1993) 279--289.

\bibitem{Vermaseren:2000nd}
J.~A.~M.~Vermaseren,
New features of FORM,
[arXiv:math-ph/0010025 [math-ph]].

\bibitem{vanRitbergen:1998pn} 
  T.~van Ritbergen, A.~N.~Schellekens, and J.~A.~M.~Vermaseren,
  Group theory factors for Feynman diagrams,
  Int.\ J.\ Mod.\ Phys.\ A {\bf 14} (1999) 41--96
  [hep-ph/9802376].

\bibitem{Ruijl:2017cxj} 
  B.~Ruijl, T.~Ueda, and J.~A.~M.~Vermaseren,
  {\sc Forcer}, a {\sc Form} program for the parametric reduction of four-loop massless propagator diagrams,
  Comput.\ Phys.\ Commun.\  {\bf 253} (2020) 107198
  [arXiv:1704.06650 [hep-ph]].

\bibitem{Moch:2017uml} 
  S.~Moch, B.~Ruijl, T.~Ueda, J.~A.~M.~Vermaseren, and A.~Vogt,
  Four-loop non-singlet splitting functions in the planar limit and beyond,
  J. High Energy Phys.\ 10 (2017) 041 
  [arXiv:1707.08315 [hep-ph]].

\bibitem{Moch:2018wjh} 
  S.~Moch, B.~Ruijl, T.~Ueda, J.~A.~M.~Vermaseren, and A.~Vogt,
  On quartic colour factors in splitting functions and the gluon cusp anomalous dimension,
  Phys.\ Lett.\ B {\bf 782} (2018) 627--632
  [arXiv:1805.09638 [hep-ph]].

\bibitem{Vermaseren:1998uu} 
  J.~A.~M.~Vermaseren,
  Harmonic sums, Mellin transforms and integrals,
  Int.\ J.\ Mod.\ Phys.\ A {\bf 14} (1999) 2037--2076
  (1999)
  [hep-ph/9806280].

\bibitem{Lipatov:1976zz}
  L.~N.~Lipatov,
  Reggeization of the Vector Meson and the Vacuum Singularity in Nonabelian Gauge Theories,
  Yad.\ Fiz.\  {\bf 23} (1976) 642--656
  [Sov.\ J.\ Nucl.\ Phys.\  {\bf 23} (1976) 338--345].

\bibitem{Kuraev:1977fs}
  E.~A.~Kuraev, L.~N.~Lipatov, and V.~S.~Fadin,
  The Pomeranchuk Singularity in Nonabelian Gauge Theories,
  Zh.\ Eksp.\ Teor.\ Fiz.\  {\bf 72} (1977) 377--389
  [Sov.\ Phys.\ JETP {\bf 45} (1977) 199--204].

\bibitem{Balitsky:1978ic}
  I.~I.~Balitsky and L.~N.~Lipatov,
  The Pomeranchuk Singularity in Quantum Chromodynamics,
  Yad.\ Fiz.\  {\bf 28} (1978) 1597--1611
  [Sov.\ J.\ Nucl.\ Phys.\  {\bf 28} (1978) 822--829].

\bibitem{Fadin:1998py}
V.~S.~Fadin and L.~N.~Lipatov,
BFKL pomeron in the next-to-leading approximation,
Phys.\ Lett.\ B \textbf{429} (1998) 127--134
[arXiv:hep-ph/9802290 [hep-ph]].

\bibitem{Kotikov:2000pm}
A.~V.~Kotikov and L.~N.~Lipatov,
NLO corrections to the BFKL equation in QCD and in supersymmetric gauge theories,
Nucl. Phys. B \textbf{582} (2000) 19--43
[arXiv:hep-ph/0004008 [hep-ph]].

\bibitem{Kirschner:1982qf}
R.~Kirschner and L.~N.~Lipatov,
  Double-logarithmic asymptotics of quark scattering amplitudes with flavor exchange,
  Phys.\ Rev.\ D {\bf 26} (1982) 1202--1205(R).

\bibitem{Kirschner:1983di}
R.~Kirschner and L.~N.~Lipatov,
Double logarithmic asymptotics and regge singularities of quark amplitudes with flavor exchange,
Nucl.\ Phys.\ B \textbf{213} (1983) 122--148.
  
\bibitem{Henn:2019swt} 
  J.~M.~Henn, G.~P.~Korchemsky, and B.~Mistlberger,
  The full four-loop cusp anomalous dimension in $\mathcal{N}=4$ super Yang-Mills and QCD,
  J. High Energy Phys.\ {\bf 04} (2020) 018 
  [arXiv:1911.10174 [hep-th]].

\bibitem{Huber:2019fxe} 
  T.~Huber, A.~von Manteuffel, E.~Panzer, R.~M.~Schabinger, and G.~Yang,
  The four-loop cusp anomalous dimension from the $\mathcal{N}=4$ Sudakov form factor,
  Phys.\ Lett.\ B {\bf 807} (2020) 135543 
  [arXiv:1912.13459 [hep-th]].

\bibitem{vonManteuffel:2020vjv} 
  A.~von Manteuffel, E.~Panzer, and R.~M.~Schabinger,
  Cusp and Collinear Anomalous Dimensions in Four-Loop QCD from Form Factors,
  Phys.\ Rev.\ Lett.\  {\bf 124} (2020) 162001 
  [arXiv:2002.04617 [hep-ph]].

\bibitem{Velizhanin:2021bdh}
V.~N.~Velizhanin,
NNNLLA BFKL pomeron eigenvalue in the planar $\mathcal{N}=4$ SYM theory,
[arXiv:2106.06527 [hep-th]].

\bibitem{Basso:2006nk}
B.~Basso and G.~P.~Korchemsky,
Anomalous dimensions of high-spin operators beyond the leading order,
Nucl.\ Phys.\ B \textbf{775} (2007) 1--30
[arXiv:hep-th/0612247 [hep-th]].

\bibitem{Dokshitzer:2005bf}
Yu.~L.~Dokshitzer, G.~Marchesini, and G.~P.~Salam,
Revisiting parton evolution and the large-$x$ limit,
Phys.\ Lett.\ B \textbf{634} (2006) 504--507
[arXiv:hep-ph/0511302 [hep-ph]].

\bibitem{Dokshitzer:2006nm}
Yu.~L.~Dokshitzer and G.~Marchesini,
$\mathcal{N}=4$ SUSY Yang--Mills: Three loops made simple(r),
Phys.\ Lett.\ B \textbf{646} (2007) 189--201
[arXiv:hep-th/0612248 [hep-th]].

\end{thebibliography}
\end{document}